\documentclass[twoside,twocolumn,9pt]{article}
\usepackage{extsizes}
\usepackage[super,sort&compress,comma]{natbib}
\usepackage[version=3]{mhchem}
\usepackage[left=1.5cm, right=1.5cm, top=1.785cm, bottom=2.0cm]{geometry}
\usepackage{balance}

\usepackage{sectsty}
\usepackage{graphicx}
\usepackage{lastpage}
\usepackage[format=plain,justification=justified,singlelinecheck=false,font={stretch=1.125,small,sf},labelfont=bf,labelsep=space]{caption}
\usepackage{float}
\usepackage{fancyhdr}
\usepackage{fnpos}
\usepackage[english]{babel}
\addto{\captionsenglish}{%
  
}
\usepackage{array}
\usepackage{droidsans}
\usepackage{charter}
\usepackage[T1]{fontenc}
\usepackage[usenames,dvipsnames]{xcolor}
\usepackage{setspace}
\usepackage[compact]{titlesec}
\usepackage{hyperref}
\usepackage{amsmath}
\usepackage{amssymb}
\usepackage{bm}
\usepackage{subcaption}

\usepackage{epstopdf}
\usepackage{siunitx}

\definecolor{cream}{RGB}{222,217,201}
\usepackage{soul}

\usepackage[normalem]{ulem}
\usepackage[textwidth=\dimexpr\textwidth-2cm\relax]{todonotes}
\makeatletter
\@mparswitchfalse%
\makeatother
\normalmarginpar 

\usepackage[T1]{fontenc}
\usepackage[utf8]{inputenc}
\usepackage{textcomp}
\usepackage{helvet}
\usepackage{microtype}

\usepackage{lastpage}
\usepackage{hyperref}

\begin{document}

\pagestyle{fancy}
\thispagestyle{plain}
\fancypagestyle{plain}{
\renewcommand{\headrulewidth}{0pt}
}

\makeFNbottom
\makeatletter
\renewcommand\LARGE{\@setfontsize\LARGE{15pt}{17}}
\renewcommand\Large{\@setfontsize\Large{12pt}{14}}
\renewcommand\large{\@setfontsize\large{10pt}{12}}
\renewcommand\footnotesize{\@setfontsize\footnotesize{7pt}{10}}
\makeatother

\renewcommand{\thefootnote}{\fnsymbol{footnote}}
\renewcommand\footnoterule{\vspace*{1pt}%
\color{cream}\hrule width 3.5in height 0.4pt \color{black}\vspace*{5pt}}
\setcounter{secnumdepth}{5}

\makeatletter
\renewcommand\@biblabel[1]{#1}
\renewcommand\@makefntext[1]%
{\noindent\makebox[0pt][r]{\@thefnmark\,}#1}
\makeatother
\renewcommand{\figurename}{\small{Fig.}~}
\sectionfont{\sffamily\Large}
\subsectionfont{\normalsize}
\subsubsectionfont{\bf}
\setstretch{1.125} 
\setlength{\skip\footins}{0.8cm}
\setlength{\footnotesep}{0.25cm}
\setlength{\jot}{10pt}
\titlespacing*{\section}{0pt}{4pt}{4pt}
\titlespacing*{\subsection}{0pt}{15pt}{1pt}

\fancyfoot{}
\fancyfoot[LO,RE]{\vspace{-7.1pt}\includegraphics[height=9pt]{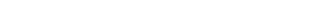}}
\fancyfoot[CO]{\vspace{-7.1pt}\hspace{13.2cm}\includegraphics{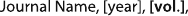}}
\fancyfoot[CE]{\vspace{-7.2pt}\hspace{-14.2cm}\includegraphics{head_foot/RF}}
\fancyfoot[RO]{\footnotesize{\sffamily{1--\pageref{LastPage} ~\textbar  \hspace{2pt}\thepage}}}
\fancyfoot[LE]{\footnotesize{\sffamily{\thepage~\textbar\hspace{3.45cm} 1--\pageref{LastPage}}}}
\fancyhead{}
\renewcommand{\headrulewidth}{0pt}
\renewcommand{\footrulewidth}{0pt}
\setlength{\arrayrulewidth}{1pt}
\setlength{\columnsep}{6.5mm}
\setlength\bibsep{1pt}

\makeatletter
\newlength{\figrulesep}
\setlength{\figrulesep}{0.5\textfloatsep}

\newcommand{\topfigrule}{\vspace*{-1pt}%
\noindent{\color{cream}\rule[-\figrulesep]{\columnwidth}{1.5pt}} }

\newcommand{\botfigrule}{\vspace*{-2pt}%
\noindent{\color{cream}\rule[\figrulesep]{\columnwidth}{1.5pt}} }

\newcommand{\dblfigrule}{\vspace*{-1pt}%
\noindent{\color{cream}\rule[-\figrulesep]{\textwidth}{1.5pt}} }

\makeatother

\twocolumn[
  \begin{@twocolumnfalse}
{{\includegraphics[height=55pt]{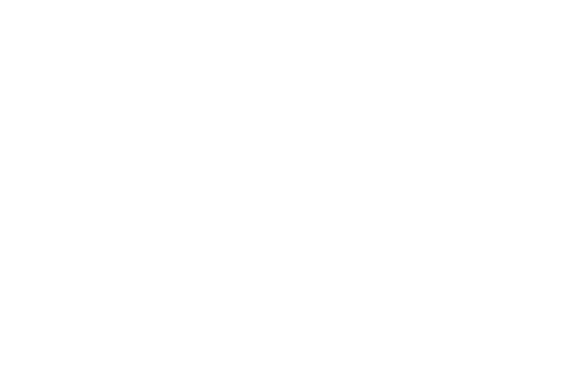}}\\[1ex]
\includegraphics[width=18.5cm]{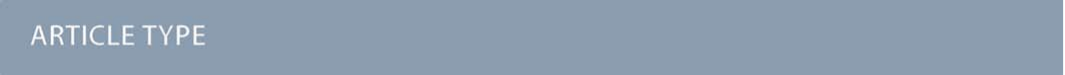}}\par
\vspace{1em}
\sffamily
\begin{tabular}{m{4.5cm} p{13.5cm} }

\includegraphics{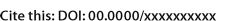} & \noindent\LARGE{\textbf{Stood-up drop to determine receding contact angles$^\dag$}} \\
\vspace{0.3cm} & \vspace{0.3cm} \\

 & \noindent\large{Diego Díaz$^{\ddag}$\textit{$^{a}$}, Aman Bhargava$^{\ddag}$\textit{$^{b}$}, Franziska Walz\textit{$^{c}$}, Azadeh Sharifi\textit{$^{c}$}, Sajjad Summaly\textit{$^{c}$}, Rüdiger Berger \textit{$^{c}$}, Michael Kappl\textit{$^{c}$},  Hans-Jürgen Butt\textit{$^{c}$}, Detlef Lohse\textit{$^{b,e}$}, Thomas Willers$^{\ast}$\textit{$^{d}$}, Vatsal Sanjay$^{\ast}$\textit{$^{b,f}$}, and Doris Vollmer$^{\ast}$\textit{$^{c}$}}\\

\includegraphics{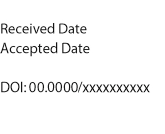} & \noindent\normalsize{The wetting behavior of drops on natural and industrial surfaces is determined by the advancing and receding contact angles. They are commonly measured by the sessile drop technique, also called goniometry, which doses liquid through a solid needle. Consequently, this method requires substantial drop volumes, long contact times, tends to be user‑dependent, and is difficult to automate. Here, we propose the stood‑up drop (SUD) technique as an alternative to measure receding contact angles. 
The method consists of depositing a liquid drop on a surface by a short liquid jet, at which it spreads radially forming a pancake‑shaped film.
Then the liquid retracts, forming a spherical cap drop shape (stood-up drop). At this quasi-equilibrium state, the contact angle ($\theta_\text{SUD}$) closely resembles the receding contact angle measured by goniometry. Our method is suitable for a wide variety of surfaces from hydrophilic to hydrophobic, overcoming typical complications of goniometry such as needle‑induced distortion of the drop shape, and it reduces user dependence. 
We delineate when the receding contact angle can be obtained by the stood‑up method using Volume‑of‑Fluid (VoF) simulations that systematically vary viscosity, contact angle, and deposited drop volume. Finally, we provide simple scaling criteria to predict when the stood‑up drop technique works.
} \\

\end{tabular}

 \end{@twocolumnfalse} \vspace{0.6cm}

  ]

\renewcommand*\rmdefault{bch}\normalfont\upshape
\rmfamily
\section*{}
\vspace{-1cm}

\footnotetext{\dag~Under consideration for publication in Soft Matter}
\footnotetext{\ddag~Contributed equally to the work}

\footnotetext{\textit{$^{a}$~Dept. of Engineering Mechanics, KTH Royal Institute of Technology, 100 44 Stockholm, Sweden}}
\footnotetext{\textit{$^{b}$~Physics of Fluids Department, Max Planck Center Twente for Complex Fluid Dynamics, Department of Science and Technology, University of Twente, P.O. Box 217, Enschede 7500 AE, Netherlands}}
\footnotetext{\textit{$^{c}$~Max Planck Institute for Polymer Research, Ackermannweg 10, 55128 Mainz, Germany}}
\footnotetext{\textit{$^{d}$~KR\"{U}SS GmBH, Wissenschaftliche Laborger\"{a}te, Borsteler Chaussee 85, 22453 Hamburg, Germany}}
\footnotetext{\textit{$^{e}$~Max Planck Institute for Dynamics and Self-Organisation, Am Fassberg 17, 37077 Göttingen, Germany}}
\footnotetext{\textit{$^{f}$~CoMPhy Lab, Department of Physics, Durham University, Science Laboratories, South Road, Durham DH1 3LE, United Kingdom}}
\footnotetext{$\ast$~Corresponding authors}





\section{Introduction}
Wetting \cite{bonn2009wetting} is ubiquitous in nature and technology, shaping phenomena such as drops rolling off plant leaves, animal skin, and insect wings, as well as applications including inkjet printing \cite{lohse2022fundamental}, coatings \cite{eral2013contact}, forensics \cite{attinger2013fluid}, disease detection \cite{yakhno2003existence}, and DNA analysis \cite{brutin2015droplet}. 
Surface wettability is characterized by measuring contact angles. Primarily, three types of contact angles exist: Young--Dupré's (equilibrium) contact angle, $\theta_Y$, advancing contact angle, $\theta_{a}$, and receding contact angle, $\theta_{r}$. Despite their relevance, measuring the different contact angles still remains challenging, in particular measuring $\theta_r$, in particular due to distortions of the drop shape by the needle. Therefore, a needle-free methood to measure the receding contact angle would be desired. 

At thermodynamic equilibrium, a stationary drop on a solid adopts $\theta_Y$ owing to the interfacial tension balance at the three phase contact line and follows Young--Dupré’s equation,\cite{young1832essay}:

\begin{equation}
    \cos{\theta_Y}= \frac{\gamma_{SG}-\gamma_{SL}}{\gamma_{LG}}
\end{equation}
where $\gamma_{SG}$, $\gamma_{SL}$ and $\gamma_{LG}$ are the solid-gas, solid-liquid and liquid-gas interfacial tensions, respectively. It is important to point out that the Young--Dupré's contact angle is a theoretical concept, often attributed to the static contact angle of sessile drops ~\cite{drelich2019contact, marmur2006soft,cwikel2010comparing, marmur2011measures, marmur2017contact}. Although the static contact angle can classify surfaces as hydrophilic, hydrophobic and superhydrophobic, it provides little information on how drops adhere to, or roll or slide over a surface.

When the three‑phase contact line moves, advancing of the liquid front yields $\theta_a$, whereas retraction over an already wetted region yields $\theta_r$. Both angles are affected by surface properties such as roughness and chemical heterogeneity \cite{mcphee2015wettability}, surface adaptation \cite{Butt2021adaptation}, and slide electrification \cite{Butt2021charging}. In practice, contact angles of resting drops lie within $[\theta_r,\theta_a]$; these bounds characterize wetting ($\theta_a$) and de‑wetting ($\theta_r$) behavior on a given surface. The receding angle is indispensable for modeling drop removal. For example, the pull‑off force per unit length is $\gamma_{LG}\left(1+\cos\theta_r\right)$ \cite{samuel2011, dupre1869theorie}. Lower $\theta_r$ promotes residual drops during withdrawal ($\theta_r < 90^\circ$), and $\theta_r$ correlates with practical adhesion in anti‑icing \cite{meuler2010relationships, golovin2016designing} and anti‑biofouling \cite{duong2015polysiloxane, yi2015ion} applications. Moreover, $\theta_r$ predicts instabilities of receding lines, the occurrence and duration of rebound \cite{antonini2013drop}, and drop–surface friction \cite{Gao2017friction}, making it central to drop dynamics on solids \cite{butt2014,Gao2017friction,Rasfriction,Aizenbergfriction}.

Standard measurements of $\theta_a$ and $\theta_r$ use sessile‑drop goniometry: a needle injects liquid into a sessile drop to increase volume (advancing), then withdraws to decrease volume (receding). Despite algorithmic advances \cite{huhtamaki2018, korhonen2013reliable, kalantarian2011development, hoorfar2004axisymmetric, srinivasan2011assessing, xu2014algorithm}, this protocol has drawbacks. The needle perturbs the free surface (capillary rise) and especially biases $\theta_r$ at small volumes; needle position can alter the contact‑line speed. Typical devices require samples of a few cm$^2$ and volumes of $\sim30$–$50~\mu$L to reach $\theta_r$, limiting heterogeneous or small specimens. The inflow/outflow cycle and setup time render the method time‑consuming and user‑dependent~\cite{jin2016replacing}. An alternative approach has also been proposed, in which a droplet is deposited through a pre-drilled hole in the substrate~\cite{KWOK199663}. Although this method can prevent needle induced effects, it represents an idealized configuration \cite{Drelich2013}, which is time consuming and can induce surface damage in fragile samples. This can alter the local topography, compromising the integrity of heterogeneous or patterned surfaces.

Here we introduce the stood‑up drop (SUD) method for receding‑angle metrology. A short liquid jet (jetting time of a few ms) deposits a thin lamella that spreads radially and then retracts to a spherical‑cap drop. After a quasi‑equilibrium is reached, the apparent contact angle $\theta_{\mathrm{SUD}}$ closely approximates $\theta_r$. The measurement completes in $\sim10$~ms and requires no needle, mitigating deposition artefacts and facilitating automation. We validate SUD by directly comparing $\theta_{\mathrm{SUD}}$ with $\theta_r$ across hydrophilic–hydrophobic substrates, and we delineate when $\theta_{\mathrm{SUD}}$ recovers $\theta_r$ using direct numerical simulations based on a Volume‑of‑Fluid (VoF) technique that systematically varies viscosity, static contact angle, and deposited volume. Finally, we summarize simple scaling criteria that predict the SUD regime and practical volume limits, and we highlight conditions under which SUD can fail (e.g., violent oscillations or detachment on highly hydrophobic surfaces). Together, these results position SUD as a fast, needle‑free alternative for robust receding‑angle characterization.

\section{Materials and experimental methods}
\subsection{Experimental setup: Stood-up drop (SUD) device}
\begin{figure}[h]
    \centering
    \includegraphics[width=\columnwidth]{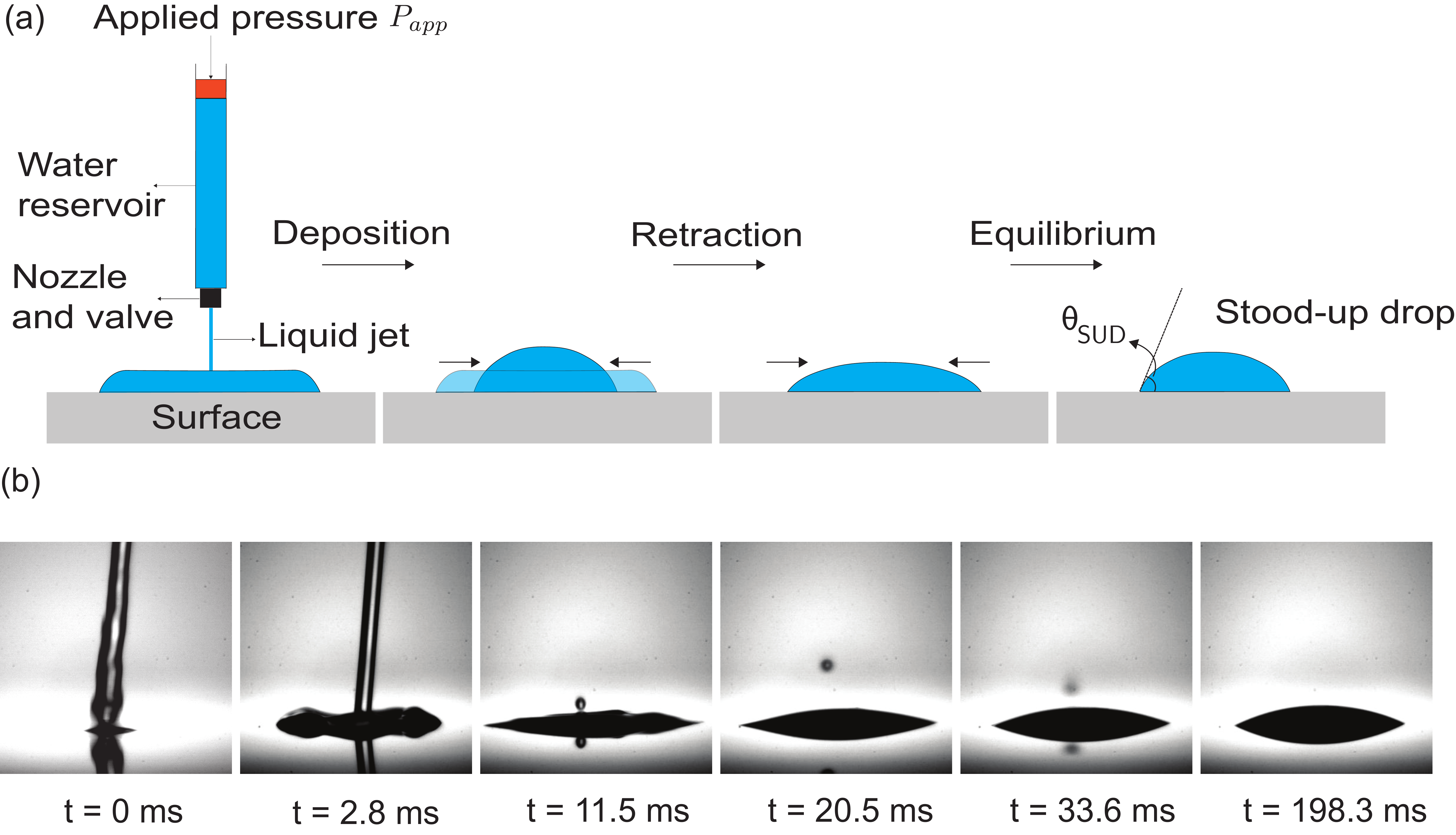}
    \caption{(a) Experimental setup for SUD technique. A liquid jet is generated at certain applied pressure $P_{app}$ for a very short jetting time $\tau$. Once the liquid impacts the sample surface, it spreads radially and then retracts. During the first retraction phase the drop can vibrate, before the contact line smoothly recedes. The liquid forms a drop with a spherical-cap shape, whose contact angle is ($\theta_\text{SUD}$). The drop profile in the first column represents the initial configuration of the drop for the numerical simulations. The configuration resembles a pancake shape, similar to that observed in experiments post-impact. The second to fourth columns sketch the retraction of the drop from an initial (lighter shade) to the quasi-equilibrium state (darker shade) as time progresses. (b) Experimental snapshots of a liquid jet impacting on a Si wafer surface at $P_{app} = 350 \;\text{mbar}$ and jetting time $\tau = 5 \;\text{ms}$. See also Video 1.}
    \label{fig:setup}
\end{figure}

To form drops from a liquid jet, a device designed by KRÜSS company was used. The device generates a liquid jet by applying a constant pressure $P_{app}$ to a liquid reservoir (Fig.~\ref{fig:setup}a), which is connected to a valve and nozzle. The distance between sample and nozzle was kept at 4 mm. The applied pressure $P_{app}$ was varied between 100 and 700 mbar.
The time at which the valve is open (jetting time $\tau$) can be controlled by software. It should not exceed a few milliseconds to ensure the formation of a thin lamella. If not stated otherwise, $\tau = 1$ ms is used. A liquid jet is generated immediately after the valve opening. After spreading and retraction of the liquid, a stationary drop is formed (Fig.~\ref{fig:setup}b). Here, the contact angle at this stage ($\theta_\text{SUD}$) is measured and compared with $\theta_{r}$ obtained by Goniometry technique. We determined $\theta_\text{SUD}$ values by a polynomial fitting method, which was designed by a python code (SI). This code determines the average of $\theta_\text{SUD}$ values from the average of contact angles at the left and right side of drops. Our results consider the $\theta_\text{SUD}$ average of the last 50 frames of each video.
 For a subset of all investigated samples no high-speed recordings were performed. In such cases, the SUD contact angle was determined using the KRÜSS ADVANCE software after drop deposition using the KRÜSS DS3251 SUD dosing unit.
 
The liquid impact process is recorded by a High-speed camera (Photron UX10) at up to 8000 fps. Simultaneous high speed-video experiments were performed from the top of the surface, to study the shape of the drop contact area and pinning of the three-phase contact line. Experiments were repeated at least three times for each surface (in some cases five times), with their wetting nature ranging from hydrophilic to hydrophobic. For each droplet, the contact angle was obtained by averaging the left and right contact angles, and then the mean and standard deviation were calculated across repeated experiments.

\subsection{Receding contact angle measurements}
Receding contact angles $\theta_{r}$ were measured by the sessile drop method/Goniometer (KRÜSS, DSA100). First, a 0.5~$\mu$L drop of Milli-Q water (18 megaohm) 
is deposited on the sample. Then, the  baseline is positioned at the contact line between drop and surface, while the needle is adjusted to be in the middle of the drop. Afterwards, the volume is increased up to 50 $\mu$L at a flow rate of 0.5~$\mu$L/s. Then, the drop is deflated at the same flow rate. A tangent fit method was used to determine receding contact angles by the ADVANCE software of the device. Experiments were repeated at least three times (up to five times in some cases) on different spots of the surface. At each time step, the mean of the left and right contact angles was taken, and the average and standard deviation of these values were then determined over the plateau region of the contact angle vs. drop base diameter plot.

\subsection{Image processing of contact angle measurements}

For automated contact angle measurements using image processing of the high speed recordings, we adapted the open source 4S-SROF toolkit to match our requirements \cite{shumaly2023deep}. The 4S-SROF toolkit effectively utilizes the OpenCV library \cite{bradski2000opencv} to manipulate the images, such as separate the drop from its background. We chose morphological transformations for noise reduction, and they proved to be superior to the median filter, guaranteeing the accuracy of our advancing angle measurements \cite{shumaly2023deep}. We calculated the advancing and receding angle using the tangent fitting method for the final 10 pixels of the drop near the substrate. The Savitzky-Golay filter \cite{press1990savitzky} was employed in specific cases to eliminate unwanted noise and enhance the smoothness of the final diagram, facilitating easier interpretation.

\subsection{Surface preparation}

Most investigated surfaces are commercially available or provided by customers from Krüss. These surfaces were cleaned by water and ethanol before usage.

\subsubsection{PFOTS coatings}

Si wafer, 1 mm thick SiO$_2$ slides  (76$\times$25$\times$5 mm$^{3}$, Thermo Fisher Scientific) and glass slides were coated with perfluorooctadecyltrichlorosilane (PFOTS). After O$_2$ plasma cleaning at 300 W for 10 min (Femto low-pressure plasma system, Diener electronic), samples were placed in a vacuum desiccator containing a vial with 0.5 mL of 1H,1H,2H,2H-perfluoro-octadecyltrichlorosilane (97\%, Sigma Aldrich). 
The desiccator was evacuated to less than 100 mbar, closed and the reaction was allowed to proceed for 30 min.

\subsubsection{PDMS brushes}
Plasma-activated silicon wafers (Si-Mat, Kaufering, Germany, plasma activation for 10 min at 300 W) were placed in a desiccator together with 1,3 dichlorotetramethyl-disiloxane (96 \%, Alfa Aesar, 50 µL in 2.150 cm$^3$) and the reaction was allowed to proceed at ambient temperature (21 °C) and ambient humidity (40 – 60 $\%$) for 3 h. Non-bonded oligomers were removed by washing the samples in toluene ($ \geq$ 9.8 $\%$, Thermo Fisher Scientific), isopropanol ($ \geq $  99.8 $\%$, Thermo Fisher Scientific), and milli-Q water for a minimum of 30 s each. 

\subsubsection{Teflon AF on ITO}

Teflon AF1600 samples were prepared by dip-coating of indium tin oxide (ITO) coated glass slides.  The slides were immersed into 1 wt. $\%$ Teflon AF1600 (Sigma Aldrich) in FC-317 (Sigma Aldrich with a speed of 90 mm/min). After being immersed for 10 s, the substrates were withdrawn from the solution at a constant speed of 10 mm/min. Finally, the coated 
substrates were annealed at $160^{\circ}$ C in a Heidolph vacuum oven for 24 h.

\section{Simulation methodology}

\subsection{Governing equations}\label{subsec:GoverningEqns}

In this section, we present the governing equations that describe the retraction dynamics of a liquid drop on a solid surface. Following the one-fluid formulation, both the drop and ambient gas are treated as a single continuum, and the interfacial boundary condition is enforced via the localized surface tension force $\boldsymbol{f_{\gamma}}$ \cite{Tryggvason_Scardovelli_Zaleski_2011}. Specifically, the drop's retraction arises from the conversion of interfacial free energy into kinetic energy, countered by viscous dissipation as captured by the Navier--Stokes equation,

\begin{equation}
    \rho \left(\frac{\partial \boldsymbol{v}}{\partial t} + \boldsymbol{\nabla\cdot}(\boldsymbol{vv})\right) = -\nabla{p} + \boldsymbol{\nabla} \cdot (2 \eta \boldsymbol{\mathcal{D}}) + \boldsymbol{f_{\gamma}} + \rho \boldsymbol{g},
    \label{eq:Navier-Stokes}
\end{equation}

\noindent with mass conservation that enforces a divergence-free velocity field $\nabla\cdot\boldsymbol{v} = 0$. Here, $\boldsymbol{v}$ and $p$ are the velocity and pressure fields respectively, and $t$ represents time. The variables $\eta$ and $\rho$ are the fluid viscosity and density, respectively. $\boldsymbol{\mathcal{D}}$ is the symmetric part of the velocity gradient tensor $(\boldsymbol{\mathcal{D}} = (\boldsymbol{\nabla}\boldsymbol{v} + (\boldsymbol{\nabla}\boldsymbol{v})^T)/2$  \cite{batchelor2000introduction}. 
Lastly, $\boldsymbol{g}$ is the acceleration due to gravity.

\subsection{Non-dimensionalization of the governing equations}

To non-dimensionalize the governing equations, we use the characteristic scales of the system: the liquid density $\rho_L$, the surface tension at the liquid-gas interface $\gamma_{LG}$, and the maximum post-impact spreading thickness of the drop $H$. These parameters define the characteristic inertiocapillary timescale ${\tau}_{\gamma}$ and the characteristic inertiocapillary velocity $v_{\gamma}$ ~\cite{Vatsal_thesis},

\begin{align}
    &{\tau}_{\gamma} = \frac{H}{v_{\gamma}} = \sqrt{\frac{\rho_L H^3}{\gamma_{LG}}},\text{ and}\\
    &v_{\gamma} = \sqrt{\frac{\gamma_{LG}}{\rho_L H}},
\end{align}

\noindent respectively. Using these characteristic scales, and using $\Tilde{x}$ to represent the dimensionless form of the respective variable $x$, we rewrite the Navier-Stokes equation Eq.~\ref{eq:Navier-Stokes} in its dimensionless form

\begin{equation}
     \left(\frac{\partial \Tilde{\boldsymbol{v}}}{\partial \Tilde{t}} + \Tilde{\nabla}\cdot(\boldsymbol{\Tilde{v}\Tilde{v}})\right) = -\Tilde{\nabla}\Tilde{{p}} + \Tilde{\nabla} \cdot (2 Oh \Tilde{\boldsymbol{\mathcal{D}}}) + \Tilde{\boldsymbol{f_{\gamma}}} + Bo \hat{\boldsymbol{g}}
     \label{eq:Dimensionless NS}
\end{equation}

\noindent where $\Tilde{p} = pH/\gamma_{LG}$ is the dimensionless pressure, $\Tilde{\boldsymbol{f_{\gamma}}} \approx \Tilde{\kappa} \Tilde{\nabla} \psi$ is the dimensionless surface tension force, where $\kappa$ is the interfacial curvature $\kappa = \boldsymbol{\nabla\cdot\hat{n}}$ with $\boldsymbol{\hat{n}}$ as the normal to the interface marked between the two fluids: liquid with $\psi = 1$ and gas with $\psi = 0$. Lastly, the Ohnesorge number and Bond number

\begin{equation}
    Oh = \frac{\eta}{\sqrt{\rho_L \gamma_{LG} H}},
    \label{eq:Ohnesorge}
\end{equation}

\begin{equation}
    Bo = \frac{\rho_L g H^2}{\gamma_{LG}}
    \label{eq:Bond}
\end{equation}

\noindent characterize the dimensionless viscosity of the retracting drop and the dimensionless gravitational acceleration, respectively. The non-dimensionalization also gives two additional control parameters, the density ratio $\rho_L/\rho_G = 1000$ and the Ohnesorge number ratio $Oh_L/Oh_G = 50$ between liquid and gas phases. These ratios remain constant throughout the paper. Additionally, the Bond number based on the formulation in Eq.~\ref{eq:Bond} is $O(10^{-3})$. At Bond numbers of this magnitude, it is expected that gravity influences the macroscale dynamics very weakly, and even more so, it is dominated by the capillary and viscous stresses near the contact line. Therefore, throughout the paper we set $Bo = 0$, with the exception of Fig.~\ref{fig:dynamics}, where the Bond number is set to be $Bo = 5.7 \times 10^{-3}$.

\subsection{Numerical setup \& domain description}

\begin{figure}
    \centering
    \includegraphics[width=1\linewidth]{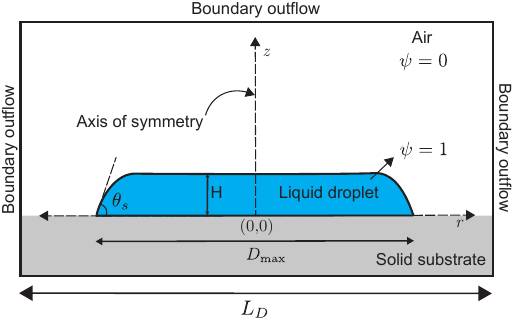}
    \caption{The initial configuration of the drop for the numerical simulations. The configuration resembles a pancake shape, similar to that observed in experiments post-impact.}
    \label{fig:drop_configuration}
\end{figure}

We perform direct numerical simulations using the Volume-of-Fluid (VoF) method, implemented using the open-source language Basilisk C ~\cite{basilisk}.
The simulations solve the governing equations presented in \S~\ref{subsec:GoverningEqns}, under the assumption of axial symmetry (see Fig.~\ref{fig:drop_configuration}). The bottom boundary of the domain represents the substrate and enforces a no-penetration condition for the normal velocity component. The tangential component is influenced by an implicit slip model, which emerges due to the interpolation of fluid properties in the VoF scheme, despite the no-slip condition being imposed ~\cite{Renardy_Renardy_Li_2001}. The implicit slip is further constrained to satisfy the prescribed contact angle $\theta$ at the first grid cell. The pressure boundary condition at this interface follows a zero-gradient condition (see Fig.~\ref{fig:drop_configuration}).

At the top and right domain boundaries, free-slip conditions are applied to the velocity field, while a Dirichlet zero condition is enforced for pressure. The domain size, denoted as $L_D$, is chosen such that these boundaries remain sufficiently distant to avoid influencing the contact line dynamics. Heuristically $L_D$ is chosen to be $D_\text{max}/(2H)+10$, where $D_\text{max}$ is the maximum spreading diameter following jet impingement. To enhance computational efficiency and accuracy, adaptive mesh refinement (AMR) is applied in regions exhibiting large velocity gradients. The refinement depth is selected to ensure a minimum of 40 grid cells across the drop thickness, i.e., $H > 40\Delta$, where $\Delta$ denotes the smallest grid cell size.

\subsection{Initial conditions}

The numerical simulations commence with the drop at rest, immediately after reaching the "pancake" configuration following impact (Fig.~\ref{fig:drop_configuration}). At this stage, the drop assumes a pancake-like shape, characterized by a thin, radially spread film with an aspect ratio $\Gamma = D_\text{max}/H$. Since the drop spreading is directly controlled by the kinetic energy provided by the jet ~\cite{Watson_1964}, we also assume that there is sufficient kinetic energy to overcome pinning and reach the pancake configuration. This assumption is justified by the drop deposition protocol in experiments where the liquid jet is turned off once a drop is created at the surface. Furthermore, the flow timescales involved in the impact and spreading process are distinct. The transition from an impacting jet to a flattened drop occurs over the kinematic timescale $D_0/V_0$ where the subsequent retractions occurs on the inertio-capillary timescale, $\tau_\gamma = \sqrt{\rho_L H^3 / \gamma_{LG}}$. Here, we focus on the retraction phase only where the influence of left-over kinetic energy from impact is assumed minimal \cite{sanjayWhenDoesImpacting2023}. 

The dimensions of the spread drop are determined by the impact conditions, with the base diameter $D_{\text{max}}$ and thickness $H$ depending on the volume of the deposited drop which depends on the nozzle flow rate $q$ and jetting time $\tau$, and the balance of inertial and capillary forces during the spreading phase. The volume of the drop $\Omega$ is related to the base diameter and thickness, upto the leading order, as
\begin{equation}
    \Omega \sim q \tau \sim D_\text{max}^2 H
\end{equation}

The jetting time is kept constant at $\tau = 1 \; \text{ms}$ during the experiments, resulting in drops with a volume of $0.5 \; \mu \text{L}$. However, we do vary the volume of the drop in our simulations, ranging from $0.25 \mu \text{L} - 3 \mu \text{L}$.
 The system is modeled under axial symmetry in a three-dimensional framework, ensuring that the retraction dynamics are captured without assuming any two-dimensional simplifications. We note that in the experiments, the drop retraction is not always axisymmetric. The initial state provides a well-defined starting point for investigating the subsequent retraction process, which is governed by surface tension and viscous effects.

\section{Results and discussion}

\subsection{SUD mechanism}

\begin{figure}
    \centering
    \includegraphics[width=\linewidth]{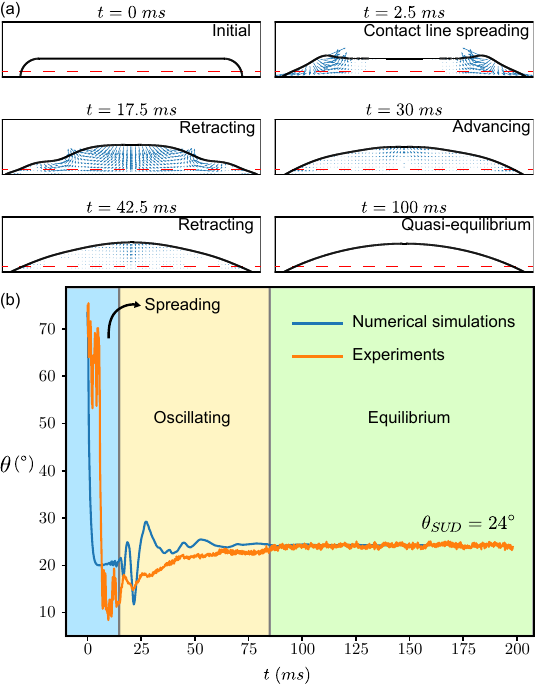}
    \caption{(a) Snapshots of a simulated drop with static contact angle $\theta_s = 27^{\circ}$ and volume of $0.5 \mu L$. The arrows represent the velocity profile inside the drop to visualize whether the drop is in the retracting or advancing phase. The dashed red line represents the height at which the contact angles are measured, in line with experimental conditions. (b) Comparison of measured contact angle from numerical simulations and experiments, as a function of time for the simulation shown in (a). $\tau = 1 \; \text{ms}$.}
    \label{fig:dynamics}
\end{figure}

Initially, a liquid jet, ejected from a nozzle at a controlled pressure (see \S~\ref{subsec:pressure}), impacts the surface with a speed of approximately $3\,\si{\meter}/\si{\second}$, as measured from the last few frames of the high-speed videos (Figs.~\ref{fig:setup}b, \ref{fig:dynamics}a, $t = 0^{+},\si{\milli\second}$).
Subsequently, the liquid spreads radially, converting kinetic energy into surface energy and viscous dissipation, forming a pancake-shaped liquid film until reaching a maximum diameter $D_\text{max}$. Analogous to impacting drops \cite{sanjayUnifyingTheoryScaling2024}, the impact and spreading duration scales with the inertiocapillary timescale. Consequently, we turn off the jet after roughly 1 ms, (Video 1).


The pancake spreading state serves as the initial condition for our simulations, assuming that the internal flow at this moment doesn't influence subsequent retraction dynamics (Fig.~\ref{fig:dynamics}a) in our simulations. For hydrophilic substrates, the contact line initially spreads to achieve its maximum spreading state. Afterwards, the pancake-shaped drop recoils, converting surface energy back into kinetic energy and viscous dissipation (Fig.~\ref{fig:dynamics}a), finally forming a static cap-spherical drop characterized by the Stood-up contact angle $\theta_\text{SUD}$ (Fig.~\ref{fig:dynamics}, $t \gtrapprox 100\,\si{\milli\second}$, Video 1).

In experiments, during spreading, the contact angle significantly decreases until reaching a minimum (Fig.~\ref{fig:dynamics}b, blue region for both experiments and simulations). The minimum is caused by the large impact pressure, resulting in the formation of a very thin lamellae. As the contact line recedes, first the contact angle oscillates (yellow region). 
The release of the stored surface energy can cause the oscillations. 
Oscillations of the contact angle level off after approximately 80 ms at a larger angle. 

The temporal development of the contact angle is well reproduced in numerical simulations (Fig.~\ref{fig:dynamics}b). Oscillations are more pronounced on hydrophobic surfaces (Fig.~S1) but they ultimately cease due to viscous dissipation, resulting in a plateau at $\theta_\text{SUD}$ (Fig.~\ref{fig:dynamics}b, blue line). Since no solid needle is required, effects on the drop shape due to the deposition protocol can be ruled out. This facilitates the fitting procedure to calculate contact angles.

\begin{figure}
    \centering
    \includegraphics[width=\columnwidth]{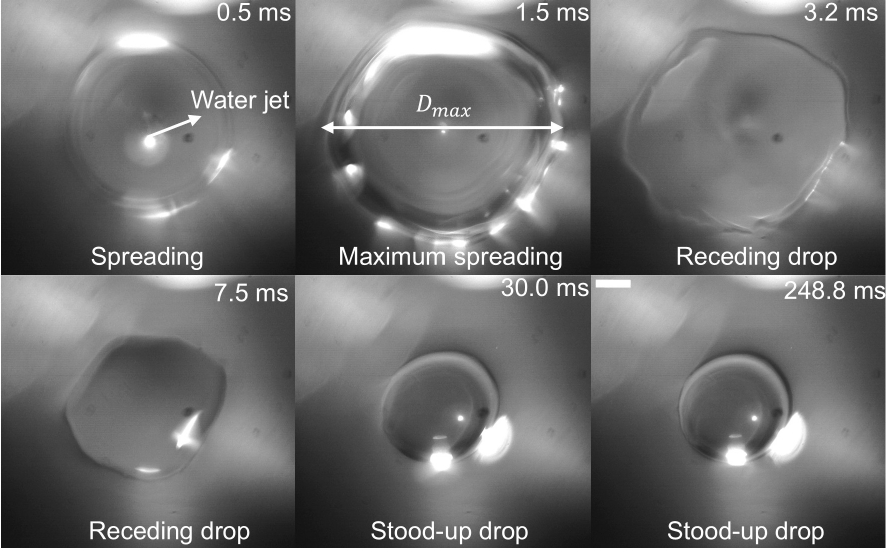}
    \caption{Top view snapshots of the impacting liquid jet on a PMMA surface. Frame rate: $4000$~fps. No pinning points are observed. Pressure: $350$~mbar. Jetting time: $\tau = 1 \; \text{ms}$. Scale bar represents $0.5$~mm. White dots correspond to specular reflections.}
    \label{fig:snapshots}
\end{figure}

To investigate possible pinning during retraction, we recorded the impacting jet from the top of the substrate using high-speed imaging. No pinning points or residual drops were observed (Fig.~\ref{fig:snapshots}). The final SUD drop area was clearly circular, Video 2. 

\begin{figure*}
    \centering
    \includegraphics[width=\textwidth]{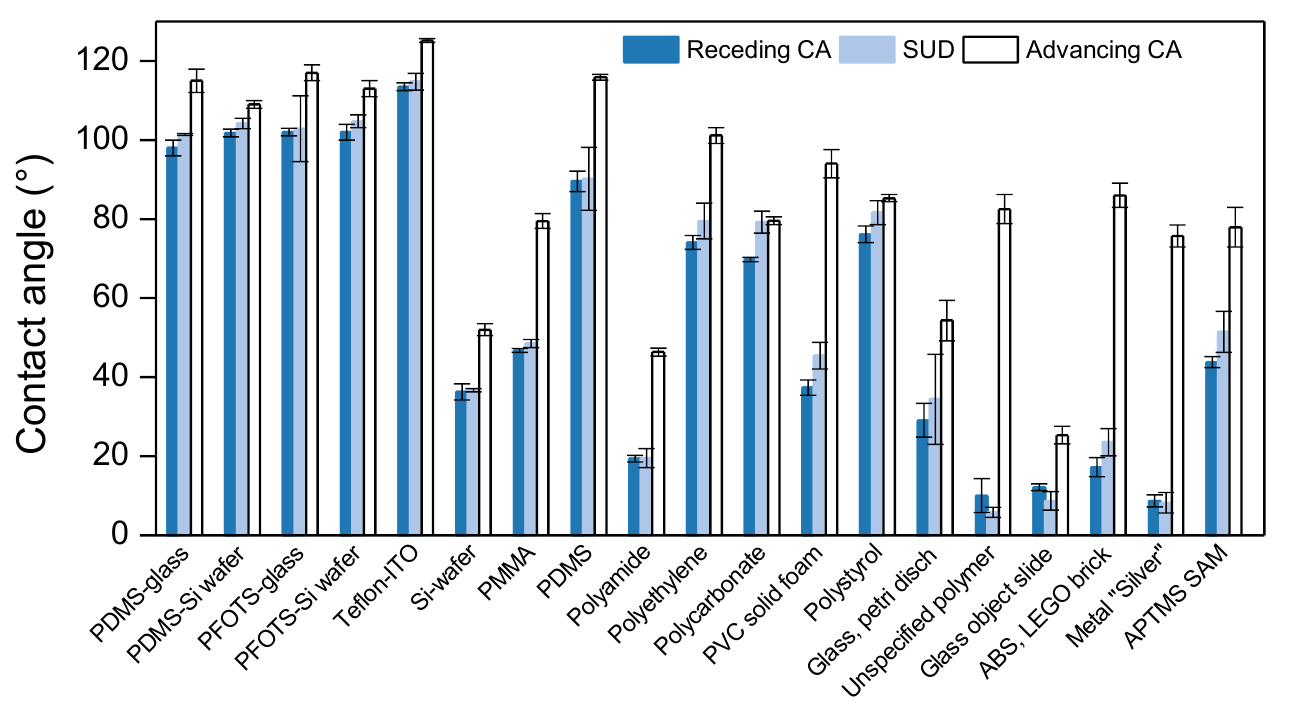}
    \caption{Comparison between the receding $\left(\theta_{r} \right)$ and advancing $\left(\theta_{a} \right)$ contact angles measured by goniometry technique (dark blue and white rectangles) and the stood-up contact angles (light blue rectangles) $\left(\theta_{\text{SUD}}\right)$. Values of $\left(\theta_{r} \right)$ and $\left(\theta_{\text{SUD}}\right)$ are very close, apart from a few samples. 
    For the following samples: PDMS/PFOTS on Si wafer and glass, Teflon on ITO, and Si wafer, $\theta_{\text{SUD}}$ was measured by the tangent fitting method described in Section 2.3, while the rest of the surfaces by the Young-Laplace fit provided by the goniometer software.}
    \label{fig:comparison}
\end{figure*}

\subsection{Further experimental results}

\subsubsection{Influence of pressure on the observed SUD contact angle}\label{subsec:pressure}

To analyze the influence of pressure on $\theta_\text{SUD}$, we generate SUD drops at three different applied pressures $P_\text{app}$ (100, 350, and 700 mbar) on PMMA. Our experiments show that $\theta_\text{SUD}$ remains largely unchanged across this pressure range (Fig.~S2). As $P_\text{app}$ increases, the liquid impacts the surface with higher speed and kinetic energy, causing the drop to spread further and reach larger maximum diameters $D_\text{max}$, thus increasing the aspect ratio $\Gamma$ -- an effect analogous to increasing the Weber number $We$ for impacting drops \cite{sanjayUnifyingTheoryScaling2024}. Nonetheless, as the drop retracts, memory effects can be excluded.
The impacting jet 
is always turned off before the retraction phase starts. As soon 
as the drop comes to rest 
the contact angle converges to a constant value of $\theta_\text{SUD}$ 
and is independent of $P_\text{app}$ \cite{PatentWillers}.

We stress that increasing $P_\text{app}$ increases both the flow rate (volume of liquid ejected per unit time during valve opening) and the maximum attainable SUD drop volume. This volume depends on $P_\text{app}$, the jetting time $\tau$, and the nozzle diameter. At the end of the retraction phase, a SUD behaves like a spherical cap with volume Fig.~\ref{fig:snapshots}:

\begin{equation}
    \Omega=\frac{\pi}{6}H_c(3r^2+H_c^2),
\end{equation}

\noindent where $H_c$ and $r$ are the height and footprint radius of the cap. However, the SUD volume variation is insignificant in our experiments due to the short $\tau$ used.
For instance, on a PMMA surface, the SUD volume increases by $\sim 0.5\,\si{\micro\litre}$
when $P_\text{app}$ changes from 350 to 700 mbar. An ejected volume that is too large can prevent the SUD state (see \S~\ref{sec:SUD-validity}).
 

In previous work \cite{PatentWillers}, we showed that the flow rate—defined by the combination of $P_\text{app}$ and the jet diameter—is the key factor determining the largest drop volume that still achieves the SUD state. If, for a given flow rate, the final drop volume is below a critical threshold, the contact line recedes after dosing, and the drop’s rest angle is the receding contact angle $\theta_r$. From experiments on six substrates \cite{PatentWillers}, we identified a phenomenological correlation between the flow rate and the maximum drop volume ensuring SUD dosing.

\subsubsection{Comparison between $\theta_\text{SUD}$ and $\theta_{r}$}

We measured the corresponding $\theta_\text{SUD}$ for different surfaces, from hydrophilic to hydrophobic and compared the values with $\theta_{r}$ obtained by the goniometer technique. Our results reveal a good agreement between $\theta_{r}$ and $\theta_\text{SUD}$ (Fig.~\ref{fig:comparison}). Agreement is slightly worse for a few samples showing particularly low contact angle hysteresis, such as polycarbonate.

Remarkably, SUD technique is suitable for a hydrophilic surface (bare glass slide), overcoming the difficulties of measuring contact angles for these surfaces by the sessile-drop method, caused by evaporation and drop pinning. As the SUD drop forms in a few miliseconds, evaporation does not affect $\theta_{\text{SUD}}$ values. Moreover, the high-speed of the liquid jet for both spreading and the first period of the retraction phase, avoid efficiently the influence of pinning points on the drop dynamics. These pinning points can prevent the contact line motion during inflation/deflation of drops using goniometry. Determining $\theta_{r}$ for a very hydrophilic glass slide using goniometry has proven challenging. However, a contact angle of $\theta_{r}= 4.5^{\circ}$ has been determined previously by capillary bridges\cite{nagy2019contact}. This is in good agreement with our $\theta_{\text{SUD}}$ values of $\sim 6^{\circ}$ (Fig. S3). Therefore, SUD method arises as a reliable alternative to determine $\theta_{r}$ for surfaces with low wettability.

For the case of hydrophobic surfaces like PFOTS and PDMS surfaces, the equilibrium state at which the SUD is formed takes longer time. This occurs due to the capillary oscillations generated during the retraction phase, when the surface energy is converted back to kinetic energy and viscous dissipation. 

Drop oscillations can lead to asymmetric contact line motion and the drops can even slide owing to this asymmetry or that of the surface orientation. The effect is more pronounced when the hydrophobicity increases, as shown for the case of Teflon AF on ITO (Fig.~S1). The SUD method fails when: (1) insufficient viscous dissipation allows drops to oscillate violently and settle in an advancing phase or (2) rapid receding motion generates enough upward momentum for the drop to detach from the surface. Both situations are common on highly hydrophobic surfaces, where bubble entrainment during rapid retraction can further complicate the dynamics \cite{bartolo2006singular, nguyen2020bubble, zhang2022impact, sanjay2025role}. Under these conditions, the SUD state is not the result of a final receding of the contact line, and thus the measured contact angles no longer correspond to $\theta_r$. Therefore, the SUD technique requires that the final motion of the contact line is a clear, dominant receding event, which is essential for reliably resembling the receding contact angle. 

A particular case arises on highly hydrophobic pillared surfaces. At sufficiently high impact pressures, the liquid jet may impale the surface, leading to a localized Wenzel state at the impact spot. If a clean receding phase is achieved at a given pressure without liquid detachment, the SUD state can still be reached. However, at higher pressures the contact line dynamics may be influenced by local liquid penetration, as previously reported for droplets impacting superhydrophobic nanostructured surfaces\cite{deng2013liquid}. This scenario would represent an additional limitation of the SUD method, with underlying physics that lies beyond the scope of the present study.

In the following section, we quantitatively analyze when the SUD method works by systematically exploring the parameter space defined by the Ohnesorge number $Oh$ and the aspect ratio $\Gamma$ at three representative static contact angles of $\theta_s = 60^\circ$ (hydrophilic), $\theta_s = 90^\circ$, and $\theta_s = 120^\circ$ (hydrophobic).

\subsection{Simulation results for the full $Oh - \Gamma$ parameter space}

\begin{figure*}[h]
    \centering
    \includegraphics[width=\linewidth]{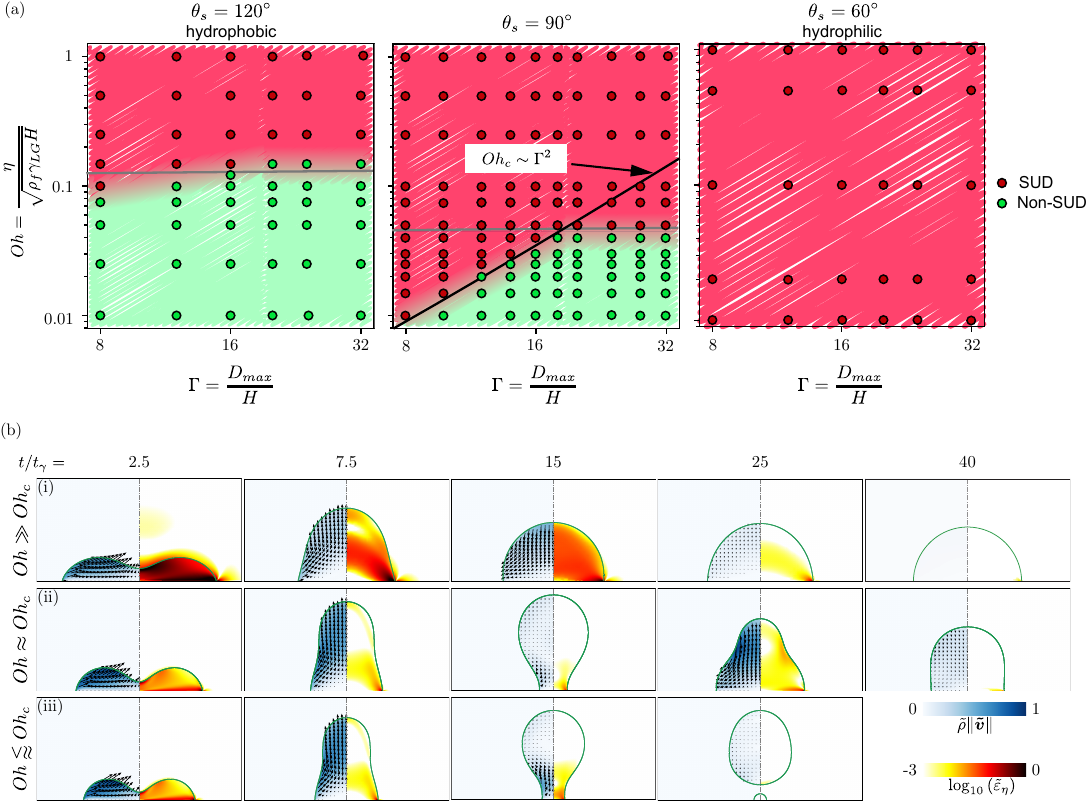}
    \caption{(a) The $Oh-\Gamma$ parameter space illustrating regions where SUD is a viable technique for $\theta = 120^\circ$ (left, hydrophobic), $\theta = 90^\circ$ (middle), and $\theta = 60^\circ$ (right, hydrophilic). (b) Numerical simulations of drop evolution for $\theta = 90^\circ$, $\Gamma = 12$ at $Oh = $ (i) $0.75 \gg Oh_c$, (ii) $0.025 \gtrapprox Oh_c$ -- a narrow neck forms at $\Tilde{t} = t/{\tau_{\gamma}} = 15$, but the drop escapes pinch off and continues surface oscillations, and (iii) $0.022 \lessapprox Oh_c$ -- a narrow neck develops which eventually pinches off. Here, $Oh_c(\Gamma,\theta)$ represents the Ohnesorge number dictating the SUD to non-SUD transition. The left half of every simulation snapshots represents the dimensionless momentum and the right half shows the dimensionless rate of viscous dissipation per unit volume normalized using the
    inertiocapillary scales, represented on a $log_{10}$ scale to differentiate the regions of maximum dissipation.. The black arrows depict the velocity vectors inside the drop.}
    \label{fig:regime_maps}
\end{figure*}

This section delineates the regions in the $Oh-\Gamma$ parameter space, for hydrophobic ($\theta_{s} > 90^{\circ}$), $\theta_s = 90^{\circ}$, and hydrophilic ($\theta_{s} < 90^{\circ}$) surfaces, where SUD is a viable technique. 
%
%
In the SUD regime, the viscous dissipation is sufficient to ensure an overdamped retracting drop that stays on the surface.
Additionally, the total volume of the drop (keeping all other material and flow properties fixed) presents additional constraints on this technique, which is discussed at the end of this section.

\subsubsection{When does the SUD technique work?}
\label{sec:SUD-validity}

For drops smaller than the gravito-capillary length ($H \ll l_c \equiv \sqrt{\gamma_{LG}/\rho_L g}$), the amount of viscous dissipation during the retraction phase largely dictates whether a drop remains attached to the surface after contact line recoiling (SUD regime), with or without oscillations or detaches from the surface (non-SUD regime)\cite{sanjayWhenDoesImpacting2023}.

Detached drops often fall back onto the surface. This leads to pronounced oscillations for drops already in a spherical cap shape, which opens the possibility of the final drop at rest to be the result of a wetting process and not of a de-wetting process. Consequently, the contact angle might not resemble the receding angle but rather a contact angle between advancing and receding angle. To systematically characterize the conditions classifying stable SUD formation versus detachment, we explored the parameter space spanned by the Ohnesorge number $Oh$ and aspect ratio $\Gamma$ across multiple static contact angles $\theta_s$, as illustrated in Fig.~\ref{fig:regime_maps}.

We observe three distinct mechanisms governing the transition between SUD and Non-SUD regimes, dependent on the static contact angle:

\begin{enumerate}
    \item For highly hydrophobic surfaces (large $\theta_s$), contact line dissipation is minimal, and the transition is primarily governed by bulk dissipation overcoming the excess surface energy—analogous to the bouncing-to-non-bouncing transition on superhydrophobic surfaces \cite{sanjayWhenDoesImpacting2023}. In this regime, the critical Ohnesorge number $Oh_c$ remains approximately constant; drops with $Oh > Oh_c$ dissipate sufficient energy to remain in the SUD regime, while those with $Oh < Oh_c$ detach.
    \item For intermediate wettability ($\theta_s \approx 90^\circ$), there is a balance between dissipation and released energy as illustrated by the numerical snapshots in Fig.\ref{fig:regime_maps}(b). For sufficiently large $Oh$, dissipation in both bulk and contact line produces SUD behavior, with the drop undergoing overdamped oscillations as it dissipates energy (Fig.~\ref{fig:regime_maps}(b-i)). Near the critical threshold $Oh_c$, drops may undergo several oscillation cycles before eventually stabilizing (Fig.~\ref{fig:regime_maps}(b-ii)). The SUD method fails when insufficient viscous dissipation allows drops to oscillate violently and drops settle in an advancing phase. Below $Oh_c$, insufficient viscous dissipation leads to detachment, with the liquid neck pinching off as shown in Fig.~\ref{fig:regime_maps}(b-iii).
    \item For hydrophilic surfaces (small $\theta_s$), contact line dissipation dominates and effectively suppresses detachment across the entire parameter range investigated -- encompassing the parameter space relevant for experimental applications. Consequently, as shown in Fig.~\ref{fig:regime_maps}(a-$\theta_s = 60^\circ$), all simulated conditions for hydrophilic surfaces result in stable SUD behavior, making the technique particularly robust for such surfaces.
\end{enumerate}

We develop a theoretical model balancing surface energy and viscous dissipation to quantify these transitions. The initial surface energy of the pancake-shaped drop at $t = 0$ is $E_\text{surf} = \gamma_{LG} A_\text{surf}$. To leading order in $\Gamma$, the surface area scales as

\begin{align}
    E_\text{surf} \sim \gamma_{LG}H^2 \Gamma^2.
\end{align}

During retraction, viscous dissipation ~\cite{Vatsal_thesis}

\begin{align}
    E_{\text{diss}} = 2\int_0^t\int_\Omega \eta\left(\boldsymbol{\mathcal{D}:\mathcal{D}}\right)d\Omega dt
\end{align}

\noindent enervates energy from the system. With velocity gradients scaling as $v_\gamma/\lambda$ (where $\lambda$ is the characteristic length scale) and dissipation occurring over volume $\Omega_\text{diss}$, we obtain

\begin{align}
    E_{\text{diss}} &\sim \eta \left(\frac{v_{\gamma}}{\lambda}\right)^2 {\tau}_{\gamma} \Omega_\text{diss} \sim \eta v_{\gamma} H \left(\frac{\Omega_\text{diss}}{\lambda^2}\right)
    \label{eq:viscous-dissipation}
\end{align}

The dominant velocity gradient arises from shear flow with $\lambda \sim H$, giving $v_\gamma/H$.

The location of viscous dissipation depends critically on the Ohnesorge number. At high $Oh$, velocity gradients develop throughout the drop immediately--analogous to the high-viscosity limit in Taylor--Culick retraction \cite{savva2009viscous,deka2020revisiting}. Here, dissipation occurs over the entire bulk volume:

\begin{equation}
    \Omega_\text{diss} \sim H^3 \Gamma^2
    \label{eq:bulk-scales}
\end{equation}

Substituting into Eq.~\ref{eq:viscous-dissipation} yields
\begin{equation}
\begin{split}
    E_\text{diss, bulk} &\sim \eta v_{\gamma} H \left(\frac{H^3 \Gamma^2}{H^2}\right) \\
    &\sim \eta V_{\gamma} H^2 \Gamma^2 \\
    &\sim \gamma_{LG} H^2 \Gamma^2 Oh
\end{split}
\end{equation}

When bulk dissipation dominates and balances the released surface energy, we get

\begin{align}
    \label{eq:Oh-constant}
    Oh_c \sim 1,
\end{align}

\noindent where $Oh_c$ is the critical Ohnesorge number for SUD behavior \cite{jha2020viscous,sanjayWhenDoesImpacting2023}.

At low $Oh$, viscous effects remain localized near the retracting contact line, with the central film region remaining nearly stationary initially. Dissipation concentrates in a boundary layer of volume \cite{Gennes_Hua_Levinson_1990,bartolo2005retraction}:

\begin{equation}
    \Omega_\text{diss} \sim \lambda^3
    \label{eq:contact-line-scales}
\end{equation}

\noindent The boundary layer thickness $\lambda$ is bounded by the drop thickness $H$ (Fig.~\ref{fig:regime_maps}). With $\lambda \sim H$, Eq.~\ref{eq:viscous-dissipation} gives

\begin{equation}
\begin{split}
    E_\text{diss, CL} &\sim \eta v_{\gamma} H \left(\frac{H^3}{H^2}\right) \\
    &\sim \eta v_{\gamma} H^2 \sim \gamma_{LG} H^2 Oh
\end{split}
\end{equation}

\noindent When contact line dissipation dominates and balances the surface energy,

\begin{align}
    \label{eq:Oh-linear}
    Oh_c \sim \Gamma^2
\end{align}

These criteria accurately demarcate the SUD and non-SUD regimes in Fig.~\ref{fig:regime_maps}(a-$\theta_s = 90^\circ$). At very large $Oh$ values, the transition becomes controlled primarily by bulk dissipation, resulting in a constant $Oh_c$ marked by the gray line for $\theta_s$ $= 120^\circ$ and $90^\circ$. Lastly, below a critical static contact angle (exemplified by $\theta_s = 60^\circ$ in Fig.~\ref{fig:regime_maps}a), the system features only SUD behavior.

A drop detaching from the surface obviously poses serious challenges for measurements using the SUD technique. Such detachment can cause the drop to exit the camera frame, entrain bubbles during pinch-off and redeposition \cite{sanjay2025role}, or significantly increase the time required to reach equilibrium. Furthermore, detached drops often fall back onto the surface, resulting in uncontrolled secondary deposition unlike the controlled jetting process discussed previously. 

Indeed, a key takeaway from the numerical simulations is that SUD technique remains viable for hydrophilic surfaces across the entire parameter space, even at extreme combinations of small $Oh$ and large $\Gamma$. For hydrophobic surfaces, however, the technique only works within the restricted range of large $Oh$ and small $\Gamma$.
Lastly, we stress that apart from the non-SUD cases discussed in this section, the technique can also fail when surfaces are superhydrophobic, causing the jet to bounce instead of depositing a drop, or when drop breakup occurs (Fig.~S4). Analysis of lift-off during jet impingement lies beyond our current scope.

\subsubsection{Other volume limitations}
In addition to directly affecting the Ohnesorge number, very small volume of the drop can cause practical issues in measuring the contact angle, such as having a limited resolution to measure the contact angle close to the surface relative to the size of the drop. On the other hand, at very large volumes, liquid jets do not reach the SUD regime as the spreading behavior is favored by the excess of volume. This prevents the onset of a receding phase. 

\section{Conclusions and Outlook}

In this work, we have introduced and validated the Stood-up drop (SUD) technique as an effective method for measuring receding contact angles. Microliter-sized water drops, formed after an impacting liquid jet spreads and recoils on a surface, exhibit a contact angle ($\theta_\text{SUD}$) that remarkably resembles the receding contact angle ($\theta_r$) measured by traditional goniometry. Our numerical simulations, performed using the Volume-of-Fluid method, corroborate our experimental findings and provide a theoretical framework for understanding this technique and its limitations.

The SUD method offers several key advantages over conventional goniometry. First, it eliminates the need for a solid needle, thereby avoiding distortion of the drop shape during measurement. Second, it dramatically reduces measurement time from minutes to milliseconds, enhancing experimental efficiency. Third, it performs exceptionally well on hydrophilic surfaces, where traditional techniques often struggle due to evaporation and pinning issues. Fourth, it requires smaller sample sizes ($\approx 0.5~\mu\text{L}$), making it suitable for testing smaller or heterogeneous surfaces. Fifth: The contact angle after SUD dosing $\theta_\text{SUD}$ describes the smallest possible contact angle of a drop at rest on the investigated surface. This in combination with the already established liquid needle dosing \cite{jin2016replacing} describing the largest possible contact angle of a drop on the investigated surface allow for the easiest way to determine contact angle hysteresis. By systematically exploring the parameter space defined by the Ohnesorge number ($Oh$) and aspect ratio ($\Gamma$), we have established clear boundaries for the applicability of this technique. For hydrophilic surfaces, the SUD method remains viable across the entire parameter space investigated. For hydrophobic surfaces, its applicability is restricted to combinations of sufficiently large $Oh$ and small $\Gamma$, where viscous dissipation prevents detachment. This framework allows researchers to determine a priori whether the SUD technique will provide reliable measurements for their specific systems. While the SUD method exhibits exceptional performance across a wide range of surfaces, it has limitations for highly hydrophobic materials. In these cases, insufficient viscous dissipation can lead to violent oscillations or detachment, preventing the formation of a stable equilibrium state dominated by receding dynamics. Additionally, volume constraints must be considered, as tiny drops present resolution challenges, while large drops deviate from the spherical cap assumption due to gravitational deformation.

Looking forward, the SUD technique opens several promising avenues for research. The temporal dynamics of the surface tension-driven receding process varies significantly between substrates (e.g., 2-3 seconds on PMMA versus approximately 30 ms on other samples). This difference remains poorly understood but might provide valuable insights into a surface's dewetting properties beyond the steady-state contact angle measurements. We have shown that the Stood-up drop technique works excellently on smooth surfaces. Future work could explore these dynamic aspects to develop a more comprehensive characterization of surface wettability, especially on more complex substrates such as sticky or textured surfaces. Moreover, the rapid and accurate determination of receding contact angles enabled by the SUD method could enhance our ability to predict contact line instabilities such as drop pinning and splitting. This has practical implications for various applications including spray coatings, self-cleaning surfaces \cite{schmidt2004contact, wisdom2013self}, anti-icing materials \cite{wang2021dynamic}, and biofouling-resistant coatings \cite{sun2025nanoparticle}, where the receding contact angle offers better correlation with practical adhesion work than the advancing contact angle \cite{de1985wetting,bonn2009wetting}. The SUD technique could also be extended to investigate more complex fluids with non-Newtonian properties \cite{jalaal2021spreading, francaElastoviscoplasticSpreadingPlastocapillarity2024} or to study the temperature dependence of receding contact angles \cite{bonn2009wetting} -- both relevant for industrial applications. Additionally, combining the SUD method with simultaneous measurements of contact line dynamics might reveal further insights into the fundamental physics of wetting and dewetting processes \cite{snoeijerMovingContactLines2013}.

%
%

\section*{Author contributions}
A.B. and D.D. contributed equally to the work. D.V., T.W. and D.D. planned the experiments. D.D. and F.W. performed the experiments, using a code written by S.S. A.S. supported with the preparation of the samples. V.S., D.L. and A.B. planned the simulations. A.B. ran the simulations. D.D., F.W. and A.B. analyzed the data. D.V., V.S., T.W., D.D. and A.B. designed the structure of the manuscript. D.D., A.B., D.V., T.W., V.S., R.B., M.K., H.-J.B. and D.L. wrote the manuscript. All authors discussed the measurements and approved the manuscript.

\section*{Conflicts of interest}
Thomas Willers is employed by KRÜSS GmBH.

\section*{Data availability}

The codes used for the numerical simulations are available at \url{https://github.com/comphy-lab/Retracting-Droplet}. The experimental data is available from the corresponding authors upon reasonable request.

\section*{Acknowledgements}
We would like to thank KRÜSS GmBH for their support with measurements. This work was carried out on the national e-infrastructure of SURFsara, a subsidiary of SURF cooperation, the collaborative ICT organization for Dutch education and research. This work was sponsored by NWO - Domain Science for the use of supercomputer facilities.

A.B., V.S., A.S., D.D., D.V., H.-J.B, and D.L. acknowledge the Max Planck–University Twente Center for Complex Fluid Dynamics for financial support. We acknowledge the ERC Advanced Grant nos. 340391-SUPRO and 740479-DDD. A.B., V.S. and D.L. acknowledge the NWO-ASML connecting industries grant for high-speed immersion lithography.

This work is supported by the German Research Foundation (DFG) within the framework Collaborative Research Centre 1194 ‘‘Interaction of Transport and Wetting Processes’’ C07N, TO2 (R.B., H.-J.B),. Open Access funding provided by the Max Planck Society.



\balance


\bibliography{rsc} 

@article{Butt2021adaptation,
author = {Butt, H.-J. and Berger, R. and Steffen, W. and Vollmer, D. and Weber, S.A.L.},
	journal = {Langmuir},
	pages = {11292--11304},
	publisher = {American Chemical Society},
	title = {Adaptive Wetting - Adaptation in Wetting},
	volume = {34},
	year = {2018}}

@article{Butt2021charging,
author = {Stetten, A.Z. and Golovko, A.S. and Weber, S.A.L. and Butt, H.-J.},
	journal = {Soft Matter},
	pages = {8667--8679},
	publisher = {Royal Society of Chemistry},
	title = {Slide electrification: charging of surfaces by moving water drops},
	volume = {15},
	year = {2019}}

@article{Rasfriction,
author = {Liimatainen, V. and Vuckovac, M. and Jokinen, V. and Sariola, V. and Hokkanen, M.J. and Zhou, Q. and Ras, R.H.A},
	journal = {Nat. Comm.},
	pages = {1798},
	publisher = {Springer},
	title = {Mapping microscale wetting variations on biological and synthetic water-repellent surfaces},
	volume = {8},
	year = {2017}}

@article{Aizenbergfriction,
author = {Daniel, D. and Timonen, J.V.I. and Li, R. and Velling, S.J. and Aizenberg, J.},
	journal = {Nature Physics},
	pages = {1020--1025},
	publisher = {Springer},
	title = {Oleoplaining droplets on lubricated surfaces},
	volume = {13},
	year = {2017}}

@article{Gao2017friction,
author = {Gao, N. and Geyer, F. Pilat, D.W. and Wooh, S. and Vollmer, D. and Butt, H.-J. and Berger, R.},
	journal = {Nature Physics},
	pages = {191–-196},
	publisher = {Springer},
	title = {How drops start sliding over solid surfaces},
	volume = {14},
	year = {2018}}

@article{savva2009viscous,
	author = {Savva, N. and Bush, J. W.},
	journal = {J. Fluid Mech.},
	pages = {211--240},
	publisher = {Cambridge University Press},
	title = {Viscous sheet retraction},
	volume = {626},
	year = {2009}}

@article{deka2020revisiting,
	author = {Deka, H. and Pierson, J.-L.},
	journal = {Phys. Rev. Fluids},
	number = {9},
	pages = {093603},
	publisher = {APS},
	title = {Revisiting the Taylor-Culick approximation. II. Retraction of a viscous sheet},
	volume = {5},
	year = {2020}}

@article{lohse2022fundamental,
	author = {Lohse, D.},
	journal = {Annu. Rev. Fluid Mech.},
	number = {1},
	pages = {349--382},
	publisher = {Annual Reviews},
	title = {Fundamental fluid dynamics challenges in inkjet printing},
	volume = {54},
	year = {2022}}

@article{francaElastoviscoplasticSpreadingPlastocapillarity2024,
	author = {Fran{\c c}a, Hugo L. and Jalaal, Maziyar and Oishi, Cassio M.},
	doi = {10.1103/PhysRevResearch.6.013226},
	journal = {Phys. Rev. Res.},
	number = {1},
	pages = {013226},
	publisher = {American Physical Society},
	shorttitle = {Elasto-Viscoplastic Spreading},
	title = {Elasto-Viscoplastic Spreading: {{From}} Plastocapillarity to Elastocapillarity},
	urldate = {2025-02-23},
	volume = {6},
	year = {2024}}

@article{jalaal2021spreading,
	author = {Jalaal, M. and Stoeber, B. and Balmforth, N. J.},
	journal = {J. Fluid Mech.},
	pages = {A21},
	publisher = {Cambridge University Press},
	title = {Spreading of viscoplastic droplets},
	volume = {914},
	year = {2021}}

@article{snoeijerMovingContactLines2013,
	author = {Snoeijer, J. H. and Andreotti, B.},
	date = {2013-01-03},
	doi = {10.1146/annurev-fluid-011212-140734},
	issn = {0066-4189, 1545-4479},
	issue = {Volume 45, 2013},
	journal = {Annu. Rev. Fluid Mech.},
	journaltitle = {Annu. Rev. Fluid Mech.},
	pages = {269--292},
	publisher = {Annual Reviews},
	shorttitle = {Moving {{Contact Lines}}},
	title = {Moving {{Contact Lines}}: {{Scales}}, {{Regimes}}, and {{Dynamical Transitions}}},
	volume = {45}}

@article{wang2021dynamic,
	author = {Wang, F. and Zhuo, Y. and He, Z. and Xiao, S. and He, J. and Zhang, Z.},
	journal = {Adv. Sci.},
	number = {21},
	pages = {2101163},
	publisher = {Wiley Online Library},
	title = {Dynamic Anti-Icing Surfaces (DAIS)},
	volume = {8},
	year = {2021}}

@article{schmidt2004contact,
	author = {Schmidt, D. L. and Brady, R. F. and Lam, K. and Schmidt, D. C. and Chaudhury, M. K.},
	journal = {Langmuir},
	number = {7},
	pages = {2830--2836},
	publisher = {ACS Publications},
	title = {Contact angle hysteresis, adhesion, and marine biofouling},
	volume = {20},
	year = {2004}}

@article{de1985wetting,
	author = {De Gennes, P.-G.},
	journal = {Rev. Mod. Phys.},
	number = {3},
	pages = {827},
	publisher = {APS},
	title = {Wetting: statics and dynamics},
	volume = {57},
	year = {1985}}

@article{wisdom2013self,
	author = {Wisdom, K. M. and Watson, J. A. and Qu, X. and Liu, F. and Watson, G. S. and Chen, C.-H.},
	journal = {Proc. Natl. Acad. Sci.},
	number = {20},
	pages = {7992--7997},
	publisher = {National Academy of Sciences},
	title = {Self-cleaning of superhydrophobic surfaces by self-propelled jumping condensate},
	volume = {110},
	year = {2013}}

@article{sun2025nanoparticle,
	author = {Sun, K. and Gizaw, Y. and Kusumaatmaja, H. and Vo{\"\i}tchovsky, K.},
	journal = {Soft Matter},
	number = {4},
	pages = {585--595},
	publisher = {Royal Society of Chemistry},
	title = {Nanoparticle adhesion at liquid interfaces},
	volume = {21},
	year = {2025}}

@article{nguyen2020bubble,
	author = {Nguyen, T. V. and Ichiki, M.},
	journal = {Microsyst. Nanoeng.},
	number = {1},
	pages = {36},
	publisher = {Nature Publishing Group UK London},
	title = {Bubble entrapment during the recoil of an impacting droplet},
	volume = {6},
	year = {2020}}

@article{jha2020viscous,
	author = {Jha, A. and Chantelot, P. and Clanet, C. and Qu{\'e}r{\'e}, D.},
	journal = {Soft Matter},
	number = {31},
	pages = {7270--7273},
	publisher = {Royal Society of Chemistry},
	title = {Viscous bouncing},
	volume = {16},
	year = {2020}}

@article{bartolo2006singular,
	author = {Bartolo, D. and Josserand, C. and Bonn, D.},
	journal = {Phys. Rev. Lett.},
	number = {12},
	pages = {124501},
	publisher = {APS},
	title = {Singular jets and bubbles in drop impact},
	volume = {96},
	year = {2006}}

@article{sanjay2025role,
	author = {Sanjay, V. and Zhang, B. and Lv, C. and Lohse, D.},
	journal = {J. Fluid Mech.},
	pages = {A6},
	publisher = {Cambridge University Press},
	title = {The role of viscosity on drop impact forces on non-wetting surfaces},
	volume = {1004},
	year = {2025}}

@article{sanjayUnifyingTheoryScaling2024,
	author = {Sanjay, V. and Lohse, D.},
	journal = {Phys. Rev. Lett.},
	number = {10},
	pages = {104003},
	publisher = {APS},
	title = {Unifying theory of scaling in drop impact: forces and maximum spreading diameter},
	volume = {134},
	year = {2025}}

@article{zhang2022impact,
	author = {Zhang, B. and Sanjay, V. and Shi, S. and Zhao, Y. and Lv, C. and Feng, X.-Q. and Lohse, D.},
	journal = {Phys. Rev. Lett.},
	number = {10},
	pages = {104501},
	publisher = {APS},
	title = {Impact forces of water drops falling on superhydrophobic surfaces},
	volume = {129},
	year = {2022}}

@article{sanjayWhenDoesImpacting2023,
	author = {Sanjay, Vatsal and Chantelot, Pierre and Lohse, Detlef},
	copyright = {Creative Commons Attribution-NonCommercial-ShareAlike 4.0 International License},
	doi = {10.1017/jfm.2023.55},
	issn = {0022-1120, 1469-7645},
	journal = {J. Fluid Mech.},
	pages = {A26},
	title = {When Does an Impacting Drop Stop Bouncing?},
	urldate = {2024-12-16},
	volume = {958},
	year = {2023}}

@inproceedings{young1832essay,
	author = {Young, Thomas},
	booktitle = {Abstracts of the Papers Printed in the Philosophical Transactions of the Royal Society of London},
	number = {1},
	organization = {The Royal Society London},
	pages = {171--172},
	title = {An essay on the cohesion of fluids},
	year = {1832}}

@article{marmur2011measures,
	author = {Marmur, A},
	journal = {The European Physical Journal Special Topics},
	number = {1},
	pages = {193},
	publisher = {Springer},
	title = {Measures of wettability of solid surfaces},
	volume = {197},
	year = {2011}}

@article{drelich2019contact,
	author = {Drelich, Jaroslaw W and Boinovich, Ludmila and Chibowski, Emil and Della Volpe, Claudio and Ho{\l}ysz, Lucyna and Marmur, Abraham and Siboni, Stefano},
	journal = {Surface Innovations},
	number = {1--2},
	pages = {3--27},
	publisher = {Thomas Telford Ltd},
	title = {Contact angles: History of over 200 years of open questions},
	volume = {8},
	year = {2019}}

@article{marmur2017contact,
	author = {Marmur, Abraham and Della Volpe, Claudio and Siboni, Stefano and Amirfazli, Alidad and Drelich, Jaroslaw W},
	journal = {Surface Innovations},
	number = {1},
	pages = {3--8},
	publisher = {Thomas Telford Ltd},
	title = {Contact angles and wettability: towards common and accurate terminology},
	volume = {5},
	year = {2017}}

@article{marmur2006soft,
	author = {Marmur, Abraham},
	journal = {Soft Matter},
	number = {1},
	pages = {12--17},
	publisher = {Royal Society of Chemistry},
	title = {Soft contact: measurement and interpretation of contact angles},
	volume = {2},
	year = {2006}}

@article{cwikel2010comparing,
	author = {Cwikel, Dory and Zhao, Qi and Liu, Chen and Su, Xueju and Marmur, Abraham},
	journal = {Langmuir},
	number = {19},
	pages = {15289--15294},
	publisher = {ACS Publications},
	title = {Comparing contact angle measurements and surface tension assessments of solid surfaces},
	volume = {26},
	year = {2010}}

@article{butt2014,
	author = {Butt, Hans-J{\"u}rgen and Roisman, Ilia V and Brinkmann, Martin and Papadopoulos, Periklis and Vollmer, Doris and Semprebon, Ciro},
	journal = {Curr. Opin. Colloid Interface Sci.},
	number = {4},
	pages = {343--354},
	publisher = {Elsevier},
	title = {Characterization of super liquid-repellent surfaces},
	volume = {19},
	year = {2014}}

@incollection{mcphee2015wettability,
	author = {McPhee, Colin and Reed, Jules and Zubizarreta, Izaskun},
	booktitle = {Developments in Petroleum Science},
	pages = {313--345},
	publisher = {Elsevier},
	title = {Wettability and wettability tests},
	volume = {64},
	year = {2015}}

@article{eral2013contact,
	author = {Eral, Husseyin B and 't Mannetje, DJCM and Oh, Jung Min},
	journal = {Colloid Polym. Sci.},
	pages = {247--260},
	publisher = {Springer},
	title = {Contact angle hysteresis: a review of fundamentals and applications},
	volume = {291},
	year = {2013}}

@book{brutin2015droplet,
	author = {Brutin, D.},
	publisher = {Academic Press},
	title = {Droplet wetting and evaporation: from pure to complex fluids},
	year = {2015}}

@article{meuler2010relationships,
	author = {Meuler, Adam J and Smith, J David and Varanasi, Kripa K and Mabry, Joseph M and McKinley, Gareth H and Cohen, Robert E},
	journal = {ACS Applied Materials \& Interfaces},
	number = {11},
	pages = {3100--3110},
	publisher = {ACS Publications},
	title = {Relationships between water wettability and ice adhesion},
	volume = {2},
	year = {2010}}

@article{golovin2016designing,
	author = {Golovin, Kevin and Kobaku, Sai PR and Lee, Duck Hyun and DiLoreto, Edward T and Mabry, Joseph M and Tuteja, Anish},
	journal = {Science Advances},
	number = {3},
	pages = {e1501496},
	publisher = {American Association for the Advancement of Science},
	title = {Designing durable icephobic surfaces},
	volume = {2},
	year = {2016}}

@article{duong2015polysiloxane,
	author = {Duong, The Hy and Briand, Jean-Fran{\c c}ois and Margaillan, Andr{\'e} and Bressy, Christine},
	journal = {ACS Applied Materials \& Interfaces},
	number = {28},
	pages = {15578--15586},
	publisher = {ACS Publications},
	title = {Polysiloxane-based block copolymers with marine bacterial anti-adhesion properties},
	volume = {7},
	year = {2015}}

@article{yi2015ion,
	author = {Yi, Zhuan and Liu, Cui-Jing and Zhu, Li-Ping and Xu, You-Yi},
	journal = {Langmuir},
	number = {29},
	pages = {7970--7979},
	publisher = {ACS Publications},
	title = {Ion exchange and antibiofouling properties of poly (ether sulfone) membranes prepared by the surface immobilization of Br{\o}nsted acidic ionic liquids via double-click reactions},
	volume = {31},
	year = {2015}}

@article{yakhno2003existence,
	author = {Yakhno, T. A. and Sedova, O. A. and Sanin, AG and Pelyushenko, AS},
	journal = {Technical Physics},
	pages = {399--403},
	publisher = {Springer},
	title = {On the existence of regular structures in liquid human blood serum (plasma) and phase transitions in the course of its drying},
	volume = {48},
	year = {2003}}

@article{huhtamaki2018,
	author = {Huhtam\"{a}ki, Tommi and Tian, Xuelin and Korhonen, Juuso T and Ras, Robin HA},
	journal = {Nature Protocols},
	number = {7},
	pages = {1521--1538},
	publisher = {Nature Publishing Group},
	title = {Surface-wetting characterization using contact-angle measurements},
	volume = {13},
	year = {2018}}

@article{korhonen2013reliable,
	author = {Korhonen, Juuso T and Huhtam\"{a}ki, Tommi and Ikkala, Olli and Ras, Robin HA},
	journal = {Langmuir},
	number = {12},
	pages = {3858--3863},
	publisher = {ACS Publications},
	title = {Reliable measurement of the receding contact angle},
	volume = {29},
	year = {2013}}

@book{kalantarian2011development,
	author = {Kalantarian, Ali},
	publisher = {University of Toronto},
	title = {Development of Axisymmetric Drop Shape Analysis-No Apex (ADSA-NA).},
	year = {2011}}

@article{hoorfar2004axisymmetric,
	author = {Hoorfar, M and Neumann, AW},
	journal = {The Journal of Adhesion},
	number = {8},
	pages = {727--743},
	publisher = {Taylor \& Francis},
	title = {Axisymmetric drop shape analysis (ADSA) for the determination of surface tension and contact angle},
	volume = {80},
	year = {2004}}

@article{srinivasan2011assessing,
	author = {Srinivasan, Siddarth and McKinley, Gareth H and Cohen, Robert E},
	journal = {Langmuir},
	number = {22},
	pages = {13582--13589},
	publisher = {ACS Publications},
	title = {Assessing the accuracy of contact angle measurements for sessile drops on liquid-repellent surfaces},
	volume = {27},
	year = {2011}}

@article{bonn2009wetting,
	author = {Bonn, D. and Eggers, J. and Indekeu, J. and Meunier, J. and Rolley, E.},
	journal = {Rev. Mod. Phys.},
	number = {2},
	pages = {739},
	publisher = {APS},
	title = {Wetting and spreading},
	volume = {81},
	year = {2009}}

@article{attinger2013fluid,
	author = {Attinger, Daniel and Moore, Craig and Donaldson, Adam and Jafari, Arian and Stone, Howard A},
	journal = {Forensic Science International},
	number = {1-3},
	pages = {375--396},
	publisher = {Elsevier},
	title = {Fluid dynamics topics in bloodstain pattern analysis: Comparative review and research opportunities},
	volume = {231},
	year = {2013}}

@article{antonini2013drop,
	author = {Antonini, Carlo and Villa, Fabio and Bernagozzi, Ilaria and Amirfazli, Alidad and Marengo, Marco},
	journal = {Langmuir},
	number = {52},
	pages = {16045--16050},
	publisher = {ACS Publications},
	title = {Drop rebound after impact: The role of the receding contact angle},
	volume = {29},
	year = {2013}}

@article{xu2014algorithm,
	author = {Xu, Z. N.},
	journal = {Rev. Sci. Instrum.},
	number = {12},
	pages = {125107},
	publisher = {AIP Publishing},
	title = {An algorithm for selecting the most accurate protocol for contact angle measurement by drop shape analysis},
	volume = {85},
	year = {2014}}

@article{samuel2011,
	author = {Samuel, Benedict and Zhao, Hong and Law, Kock-Yee},
	journal = {The Journal of Physical Chemistry C},
	number = {30},
	pages = {14852--14861},
	publisher = {ACS Publications},
	title = {Study of wetting and adhesion interactions between water and various polymer and superhydrophobic surfaces},
	volume = {115},
	year = {2011}}

@book{dupre1869theorie,
	author = {Dupr{\'e}, Athanase and Dupr{\'e}, Paul},
	publisher = {Gauthier-Villars},
	title = {Th{\'e}orie M{\'e}canique de la Chaleur},
	year = {1869}}

@article{nagy2019contact,
	author = {Nagy, Norbert},
	journal = {Langmuir},
	number = {15},
	pages = {5202--5212},
	publisher = {ACS Publications},
	title = {Contact angle determination on hydrophilic and superhydrophilic surfaces by using r--$\theta$-type capillary bridges},
	volume = {35},
	year = {2019}}

@article{jin2016replacing,
	author = {Jin, M. and Sanedrin, R. and Frese, D. and Scheithauer, C. and Willers, T.},
	journal = {Colloid and Polymer Science},
	number = {4},
	pages = {657--665},
	publisher = {Springer},
	title = {Replacing the solid needle by a liquid one when measuring static and advancing contact angles},
	volume = {294},
	year = {2016}}

@article{bartolo2005retraction,
	author = {Bartolo, D. and Josserand, C. and Bonn, D.},
	journal = {J. Fluid Mech.},
	pages = {329--338},
	publisher = {Cambridge University Press},
	title = {Retraction dynamics of aqueous drops upon impact on non-wetting surfaces},
	volume = {545},
	year = {2005}}

@article{shumaly2023deep,
	author = {Shumaly, S. and Darvish, F. and Li, X. and Saal, A. and Hinduja, C. and Steffen, W. and Kukharenko, O. and Butt, H.-J. and Berger, R.},
	journal = {Langmuir},
	number = {3},
	pages = {1111--1122},
	publisher = {ACS Publications},
	title = {Deep Learning to Analyze Sliding Drops},
	volume = {39},
	year = {2023}}

@article{bradski2000opencv,
	author = {Bradski, G.},
	journal = {Dr. Dobb's Journal: Software Tools for the Professional Programmer},
	number = {11},
	pages = {120--123},
	publisher = {Miller Freeman Inc.},
	title = {The openCV library.},
	volume = {25},
	year = {2000}}

@article{press1990savitzky,
	author = {Press, W. H. and Teukolsky, S. A.},
	journal = {Comput. Phys.},
	number = {6},
	pages = {669--672},
	publisher = {American Institute of Physics},
	title = {Savitzky-Golay smoothing filters},
	volume = {4},
	year = {1990}}

@misc{basilisk,
	author = {Popinet, S. and {Collaborators}},
	note = {Last accessed: July 22, 2025},
	title = {Basilisk},
	url = {http://basilisk.fr},
	year = {2013-2025}}

@book{Tryggvason_Scardovelli_Zaleski_2011,
	author = {Tryggvason, G. and Scardovelli, R. and Zaleski, S.},
	place = {Cambridge},
	publisher = {Cambridge University Press},
	title = {Direct Numerical Simulations of Gas--Liquid Multiphase Flows},
	year = {2011}}

@article{Gennes_Hua_Levinson_1990,
	author = {Gennes, P. G. De and Hua, X. and Levinson, P.},
	doi = {10.1017/S0022112090001859},
	journal = {J. Fluid Mech.},
	pages = {55--63},
	title = {Dynamics of wetting: local contact angles},
	volume = {212},
	year = {1990}}

@article{PatentWillers,
	author = {Willers, T. and Oetjen K.},
	journal = {Europan Patent EP4109072B1},
	title = {METHOD AND APPARATUS FOR MEASURING A RECEDING CONTACT ANGLE},
	year = {2022}}

@phdthesis{Vatsal_thesis,
title = "Viscous Free-Surface Flows",
author = "Vatsal Sanjay",
year = "2022",
month = jul,
day = "15",
doi = "10.3990/1.9789036554077",
language = "English",
isbn = "978-90-365-5407-7",
publisher = "University of Twente",
address = "Netherlands",
type = "PhD Thesis - Research UT, graduation UT",
school = "University of Twente",
}

@article{Renardy_Renardy_Li_2001,
title = {Numerical Simulation of Moving Contact Line Problems Using a Volume-of-Fluid Method},
journal = {Journal of Computational Physics},
volume = {171},
number = {1},
pages = {243-263},
year = {2001},
issn = {0021-9991},
doi = {https://doi.org/10.1006/jcph.2001.6785},
url = {https://www.sciencedirect.com/science/article/pii/S0021999101967853},
author = {Michael Renardy and Yuriko Renardy and Jie Li}
}

@book{batchelor2000introduction,
  title={An introduction to fluid dynamics},
  author={Batchelor, George Keith},
  year={2000},
  publisher={Cambridge University Press}
}

@article{Watson_1964, title={The radial spread of a liquid jet over a horizontal plane}, volume={20}, DOI={10.1017/S0022112064001367}, number={3}, journal={Journal of Fluid Mechanics}, author={Watson, E. J.}, year={1964}, pages={481–499}}

@article{Drelich2013,
    author = {Drelich, Jaroslaw},
    title = {Guidelines to measurements of reproducible contact angles using a sessile-drop technique},
    journal = {Surface Innovations},
    volume = {1},
    number = {4},
    pages = {248-254},
    year = {2013},
    month = {12},
    issn = {2050-6252},
    
}

@article{KWOK199663,
title = {Low-rate dynamic and static contact angles and the determination of solid surface tensions},
journal = {Colloids and Surfaces A: Physicochemical and Engineering Aspects},
volume = {116},
number = {1},
pages = {63-77},
year = {1996},
note = {A collection of papers presented at the Wetting and Interfacial Phenomena Symposium at the 69th Annual Colloid and Surface Science Symposium},
issn = {0927-7757},

}

@article{deng2013liquid,
  title={Liquid drops impacting superamphiphobic coatings},
  author={Deng, Xu and Schellenberger, Frank and Papadopoulos, Periklis and Vollmer, Doris and Butt, Hans-Jurgen},
  journal={Langmuir},
  volume={29},
  number={25},
  pages={7847--7856},
  year={2013},
  publisher={ACS Publications}
}
\bibliographystyle{rsc} 

\end{document}